\documentstyle[aps,epsfig,epsf,amsbsy,12pt]{revtex}
\begin{document}

\renewcommand{\baselinestretch}{1}
\newcommand{\be}{\begin{equation}}
\newcommand{\ee}{\end{equation}}

\title{First-order chiral phase transition may naturally
lead to the ``quenched'' initial condition and strong soft-pion fields}

\author{O. Scavenius$^{1,2}$, A. Dumitru$^1$}
\address{$^1$Physics Department, Yale University,
P.O.\ Box 208124, New Haven, CT 06520, USA\\
$^2$ The Niels Bohr Institute, Blegdamsvej 17, DK-2100 Copenhagen {\O},
Denmark\\}
\date{\today}

\maketitle

\begin{abstract}
We propose a novel mechanism for DCC formation in a first-order 
chiral phase transition. In this case the effective potential for the chiral 
order parameter has a local minimum at $\Phi\sim 0$ in which the chiral field
can be ``trapped''. If the expansion is sufficiently fast 
 a bubble of disoriented chiral field can emerge and decouple
 from the rest of the fireball. 
The bubble may overshoot the
mixed phase and subsequently supercool until the barrier
 disappears,
when the potential resembles that at $T=0$. This situation corresponds to the
 initial condition realized in
 a ``quench''.
Thus, the subsequent alignment in the vacuum direction leads to
 strong amplification of low momentum modes of the pion field.
We propose that these DCCs could accompany the previously suggested baryon 
rapidity
fluctuations. 
\end{abstract}

Relativistic heavy-ion collisions might offer the interesting opportunity to
study chiral symmetry restoration at non-zero temperature and density, which
could possibly lead to the formation of domains of disoriented chiral
condensate (DCC)~\cite{Anselm:1989pk,Blaizot:1992at,RW,Dyn,Ove}. 
The strongest amplification of the pion field is obtained for  
the so-called ``quenched'' initial condition~\cite{RW}. It is assumed that
the heatbath is removed instantaneously after restoration of chiral symmetry.

However, dynamical simulations~\cite{Dyn,Ove}
show that the ``quench'' does not emerge naturally in a heavy ion collision,
if the chiral phase transition is second-order or a smooth crossover. 
In this letter, we instead propose a new approach to obtain the ``quenched''
initial conditions naturally in the presence of a first-order phase transition.

It has been argued \cite{Stephanov:1999zu} that the phase
transition for 
two massless quarks at baryon-chemical potential $\mu=0$ is second-order
which then becomes a smooth crossover for small quark masses.
On the other hand, a first-order phase transition is predicted for small
 temperatures and large $\mu$. If, indeed, there is a smooth crossover for 
$\mu=0$ and
 non-zero  $T$, and a first-order transition for small $T$ and non-zero $\mu$, 
then the first-order phase transition line in the $(\mu,T)$ plane must end 
in a second-order critical point. This point is predicted to be at $T\sim
100$~MeV
 and $\mu\sim 600$~MeV. 
However, some lattice QCD results indicate
a first-order transition even at vanishing baryon-chemical
potential~\cite{lattice}.

Such temperatures and baryon-chemical potentials can be reached in
the central region of heavy-ion collisions in the forthcoming
Pb(40~AGeV)+Pb experiments at the CERN-SPS~\cite{lvbr}, and in the
fragmentation regions of more
energetic collisions at the CERN-SPS, BNL-RHIC, and CERN-LHC
($\sqrt{s}\simeq 20$, 200, 5000~AGeV)~\cite{muT}.
Furthermore, fluctuations in individual events can also provide rapidity
bins with significantly higher $\mu$ and lower $T$ than on
average~\cite{EbE,andy,Mishustin,SG2}.
In any case, the dynamical scenario for DCC formation described in this
Letter applies to the case of a first-order chiral phase transition,
and is qualitatively independent of the value of
$\mu$. Our calculation described below has been performed at $\mu=0$,
and the parameters of the Lagrangian (in particular the coupling of the
chiral field to the heat-bath) have been {\em chosen} such as to
yield a first-order phase transition. It can be viewed as a 
representative
example to illustrate the idea; the basic mechanism works equally well 
also
at non-zero $\mu$.

In case of a first-order transition,
the thermodynamical potential as function of the order parameter $\Phi$ 
exhibits a local minimum at $\Phi=0$ both in the chirally restored phase as 
well as in the mixed phase~\cite{andy,Mishustin,Agnes}.
In high-energy heavy-ion collisions the expansion rate of the locally
comoving three-volume element can become large~\cite{adrian}.
This opens the possibility that the system can break up into smaller
droplets~\cite{Mishustin,Csernai:1995zn},
which might not be able to follow an
adiabatic expansion. Instead, a bubble can ``overshoot'' the phase
boundary~\cite{andy} into the low-density broken phase.
The chiral field in the bubble is coherent if the bubble-radius
is on the order of the coherence length or smaller. Close to a first-order
phase boundary, the chiral field is light and this coherence length is 
expected to be large.

Suppose the bubble is created in the restored phase, close to the first-order
phase boundary.
Following~\cite{andy} we suppose that the chiral field within the
bubble is trapped in the $\Phi\sim0$ local minimum of the potential
(chiral symmetry is still restored but the field oscillates in the false
direction), while in previous work~\cite{Mishustin,Vischer} it was assumed
that chiral symmetry breaking had already occured in the bubble.

The preceding expansion will lead to a velocity profile in the bubble.
In other words,
the bubble will exhibit a Hubble-like expansion with a very large
expansion rate ~\cite{Mishustin,adrian}, and supercool.
During the period of supercooling
the chiral field oscillates coherently within the
whole bubble, while its energy dissipates partly due to
friction (coupling to the heat-bath \cite{friction}). 
The local minimum in the effective potential persists until the droplet
reaches the spinodal line where the potential is close to that at $T=0$
(one single minimum).  The moment where the coherent field ``leaps'' over the
barrier depends on the barrier height and the fluctuation. In a quasi-static
situation the field would tunnel to the global minimum. However, in a high
energy heavy ion collision the local expansion rate is so large (roughly
$10^{20}$ times larger than that of the universe
at spontaneous chiral symmetry breaking \cite{adrian}) that it
is reasonable to assume that tunneling has no time to occur.
\begin{figure}[htp]
\centerline{\hbox{\epsfig{figure=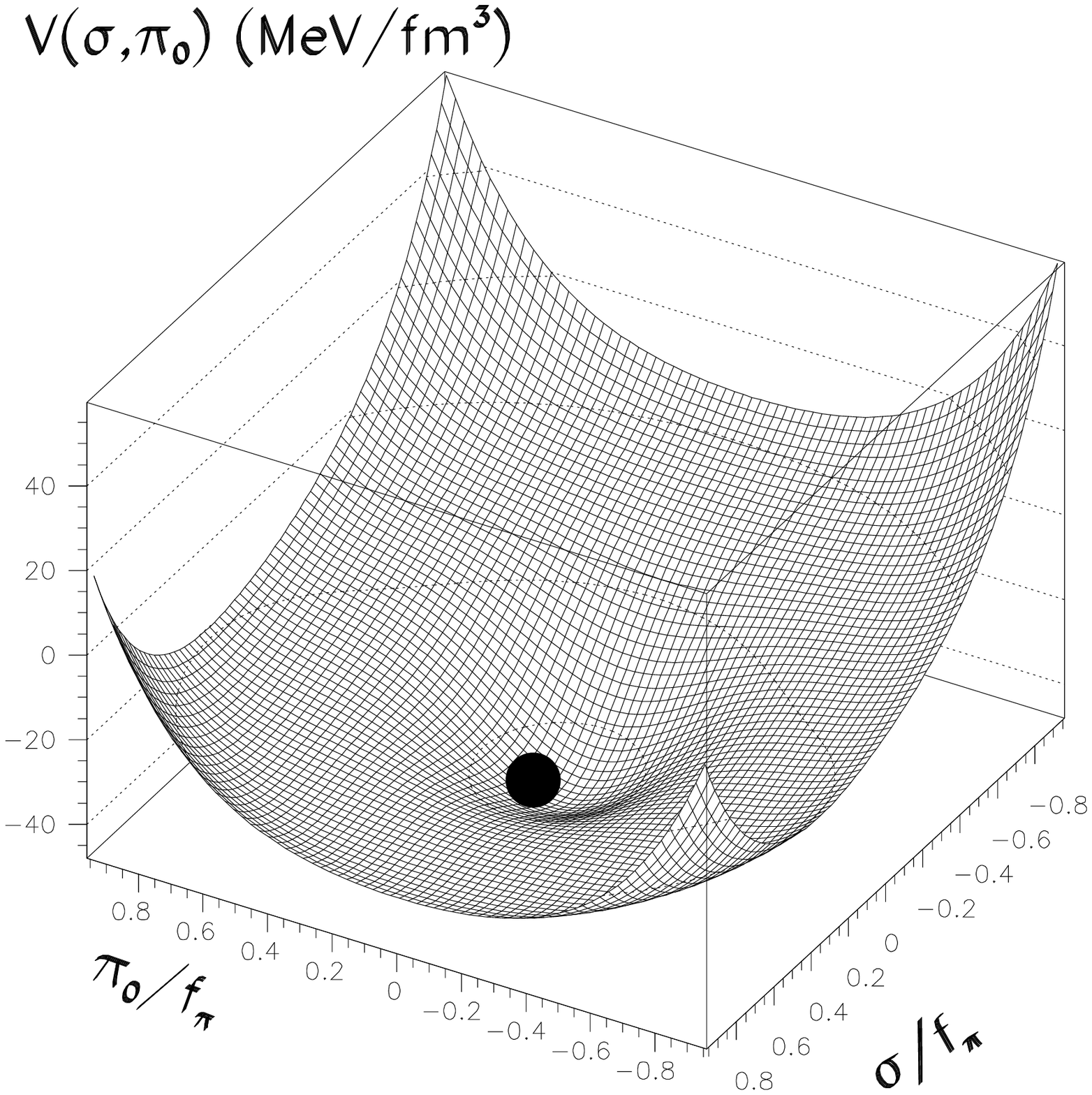,height=8cm}
\epsfig{figure=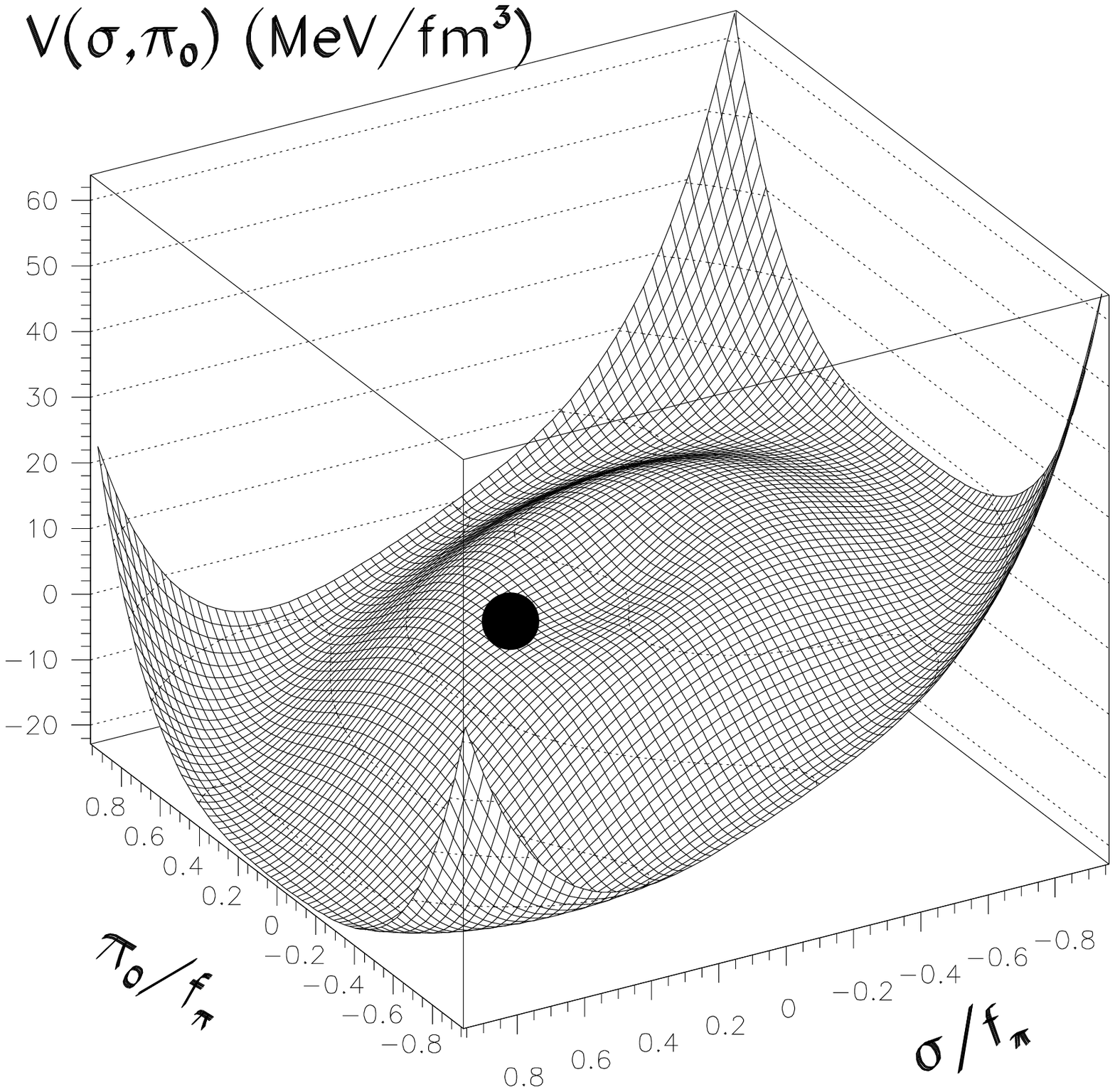,height=8cm}}}
\caption{Finite temperature effective potential (i.e.\ the grand canonical
potential) within the linear
$\sigma$-model at $T=130$~MeV (left) and $T=100$~MeV (right). The black
dot depicts the average chiral field at the corresponding time.}
\label{potstrong}
\end{figure}  

A second possibility for the chiral field to be kicked over the barrier
to the global minimum at $\Phi\neq0$ is via a thermal fluctuation.
Estimates within homogeneous nucleation theory~\cite{HNT} show that the
time-scale for nucleation of critically-sized bubbles with $\Phi\neq0$ is
about the same as that needed to reach the spinodal region ($\approx2$~fm/c
in our calculation below). Finite-size effects delay bubble nucleation
even further~\cite{FS_HNT}: if our entire decoupled bubble has the size
of the critically-sized bubble in homogeneous nucleation theory, it can
not convert to the broken phase via homogeneous nucleation. Instead, it must
supercool until it reaches the spinodal instability (when the potential
resembles that at $T=0$), cf.\ also~\cite{andy}.
Thus, the most
favorable scenario for amplification of the low-momentum modes of the pion
field, i.e.\ the ``quenched'' inital condition, is automatically realized in a
natural way if the chiral phase transition is first-order. 

Similar to previous studies of DCC formation our scenario also requires
a rapid evolution out of equilibrium. However,
the required 10\%-20\% supercooling appear much more moderate than the
instantaneous removal of the heat-bath at $T\sim T_C$, as is necessary
to obtain the ``quenched'' initial conditions in a smooth crossover~\cite{RW}.

After the true vacuum is reached, the coherent chiral field will eventually
decay into pions due to residual interactions~\cite{Blaizot:1992at,DCCdecay}.
If, at this stage, the heat-bath is still very hot and dense,
scattering of the DCC-pions with particles from the heat-bath
will randomize the isospin orientation and spread the momentum
distribution~\cite{marcus}.
However, in case of a first-order transition with supercooling,
the DCC decays at a much lower temperature and density of the heat-bath.
Therefore, it might be more feasible to detect the DCC-pions.
These pions will be blue-shifted, though, according to the velocity of
the bubble from which they emerged.

Above we discussed the mechanism for DCC formation within a decoupled bubble.
In principle, however, the same idea can be applied to the entire fireball,
thus assuming that it supercools and reaches the spinodal
instability as a whole.
Nevertheless, we have chosen to describe how our picture
works in a smaller droplet.
%This is, however, not necessary.

To illustrate the above idea, we applied the linear
$\sigma$-model coupled to a heat-bath~\cite{Ove,Csernai:1995zn}. 
The Lagrangian of the linear sigma model with quark degrees of freedom reads
\begin{equation}
{\cal L}=\overline{q}[i\gamma ^{\mu}\partial _{\mu}-g(\sigma +i\gamma _{5}
\vec{\tau} \cdot \vec{\pi} )]q
+ \frac{1}{2}(\partial _{\mu}\sigma \partial ^{\mu}\sigma + \partial _{\mu}
\vec{\pi} \partial ^{\mu}\vec{\pi} )-U(\sigma ,\vec{\pi}),
\label{sigma}
\end{equation}
where the zero temperature potential is
\begin{equation}
U(\sigma ,\vec{\pi} )=\frac{\lambda ^{2}}{4}(\sigma ^{2}+\vec{\pi} ^{2} -
{\it v}^{2})^{2}-H\sigma.
\end{equation}
Here $q$ is the light quark field $q=(u,d)$. The
scalar field $\sigma$ and the pion field $\vec{\pi} 
=(\pi _{1},\pi _{2},\pi _{3})$ 
together form a chiral field $\Phi =(\sigma,\vec{\pi})$. This Lagrangian is 
invariant under chiral $SU_{L}(2) \otimes SU_{R}(2)$ transformations if the 
explicit symmetry breaking term $H\sigma $ is zero. The parameters of the 
Lagrangian are usually chosen such that the chiral symmetry is spontaneously 
broken in the vacuum and the expectation values of the meson fields are 
$\langle\sigma\rangle ={\it f}_{\pi}$ and $\langle\vec{\pi}\rangle =0$, where
${\it f}_{\pi}=93$ MeV is the pion decay constant. The constant $H$ is fixed
by the PCAC relation which gives $H=f_{\pi}m_{\pi}^{2}$, where
$m_{\pi}=138$~MeV is the pion mass. Then one finds $v^{2}=f^{2}_{\pi}-
\frac{m^{2}_{\pi}}{\lambda ^{2}}$. The sigma mass, $m^2_\sigma=2
\lambda^{2}f^{2}_{\pi}+m^{2}_{\pi}$, which we set to 600 MeV yields 
$\lambda^{2}\approx20$. 
The coupling to the heat-bath, $g=5.5$, is chosen such as to obtain the 
potential
shown in Fig.~\ref{potstrong}. It corresponds to a first-order chiral phase 
transition. 
Note that this large value for $g$ results in a constituent 
quark mass at $T=0$ of $m_q=gf_\pi\sim512$~MeV. Thus, the nucleon mass is too 
large.
However, this is not relevant for our present considerations.
A stronger coupling leads to a stronger first-order phase transition with a more
pronounced barrier and even larger constituent quark mass; on the other hand,
$g=3.3$ fits the nucleon mass in the vacuum but results in a smooth 
crossover~\cite{Ove}.

The Euler-Lagrange equations of motion for the fields,
\begin{displaymath}
\partial_{\mu}\partial^{\mu}\sigma+\lambda^2[\sigma^2+\vec{\pi}^2-v^2]\sigma-H=
-g\rho_s,
\end{displaymath}
\be \label{EulerLagrange}
\partial_{\mu}\partial^{\mu}\vec{\pi}+\lambda^2[\sigma^2+\vec{\pi}^2-v^2]\vec{\pi
}=-g\vec{\rho}_{ps},
\ee
are solved self-consistently through the effective quark and antiquark mass, 
$m_q=g\sqrt{\sigma^2+\vec{\pi}^2}$, with the continuity equation for the 
energy-momentum
tensor of the heat-bath, which is constituted by the quarks:
\be \label{EMTensor}
\partial _{\mu}T^{\mu\nu}+\rho_{s}\partial ^{\nu}m_{q}=0,
%\partial_\mu T^{\mu\nu} = ...
\ee
where $\rho_s$ and $\vec{\rho}_{ps}$ are the scalar and pseudoscalar densities,
respectively.
To solve eqs.~(\ref{EulerLagrange}) we employ a second-order leap-frog 
algorithm,
while eqs.~(\ref{EMTensor}) are solved with the RHLLE algorithm~\cite{Dirk}, 
assuming
that $T^{\mu\nu}$ is that of an ideal fluid in local thermodynamical 
equilibrium.
For more details please refer to~\cite{Ove}.

\begin{figure}[htp]
\centerline{\hbox{\epsfig{figure=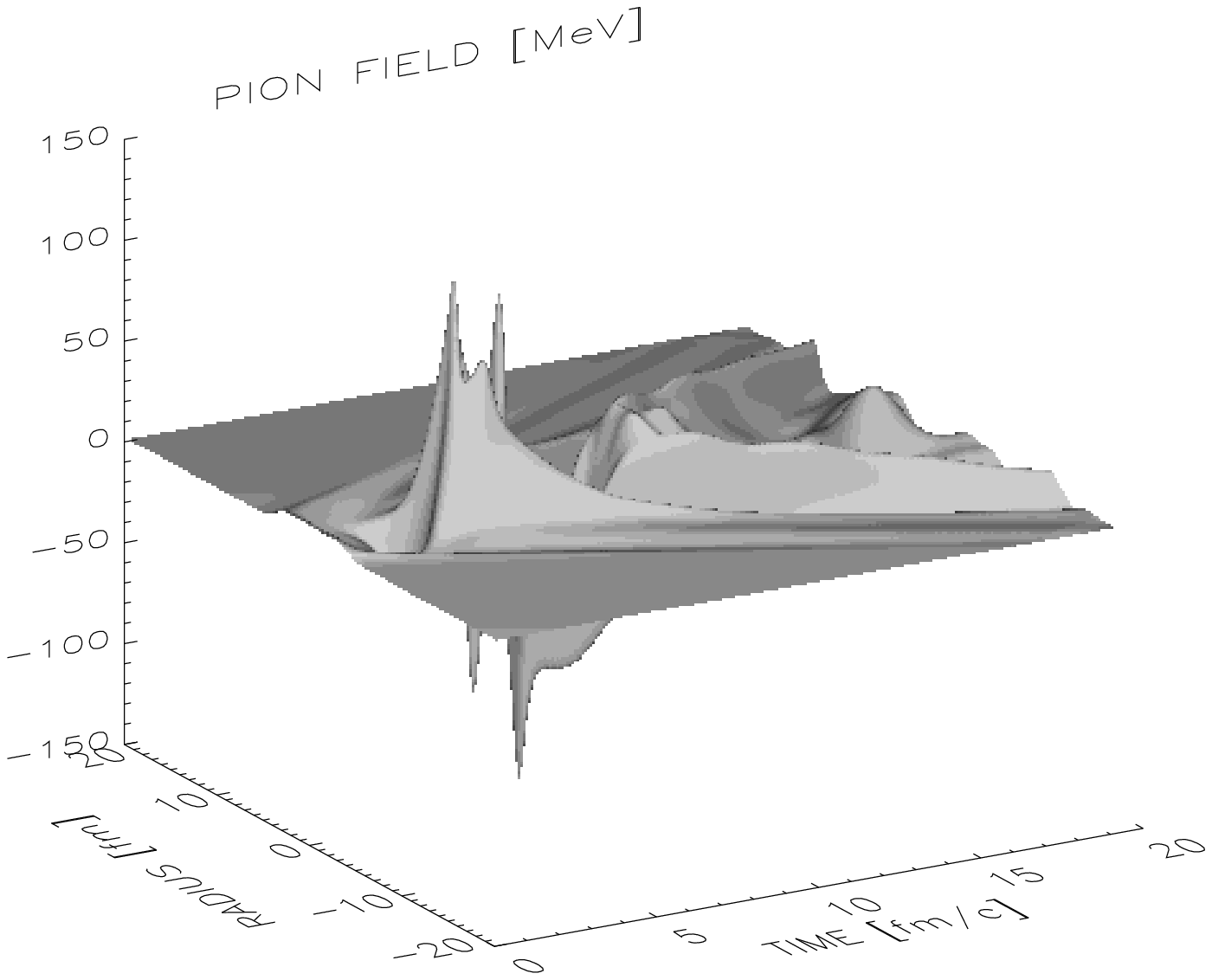,height=7.5cm} \hspace{-2.5cm}
\epsfig{figure=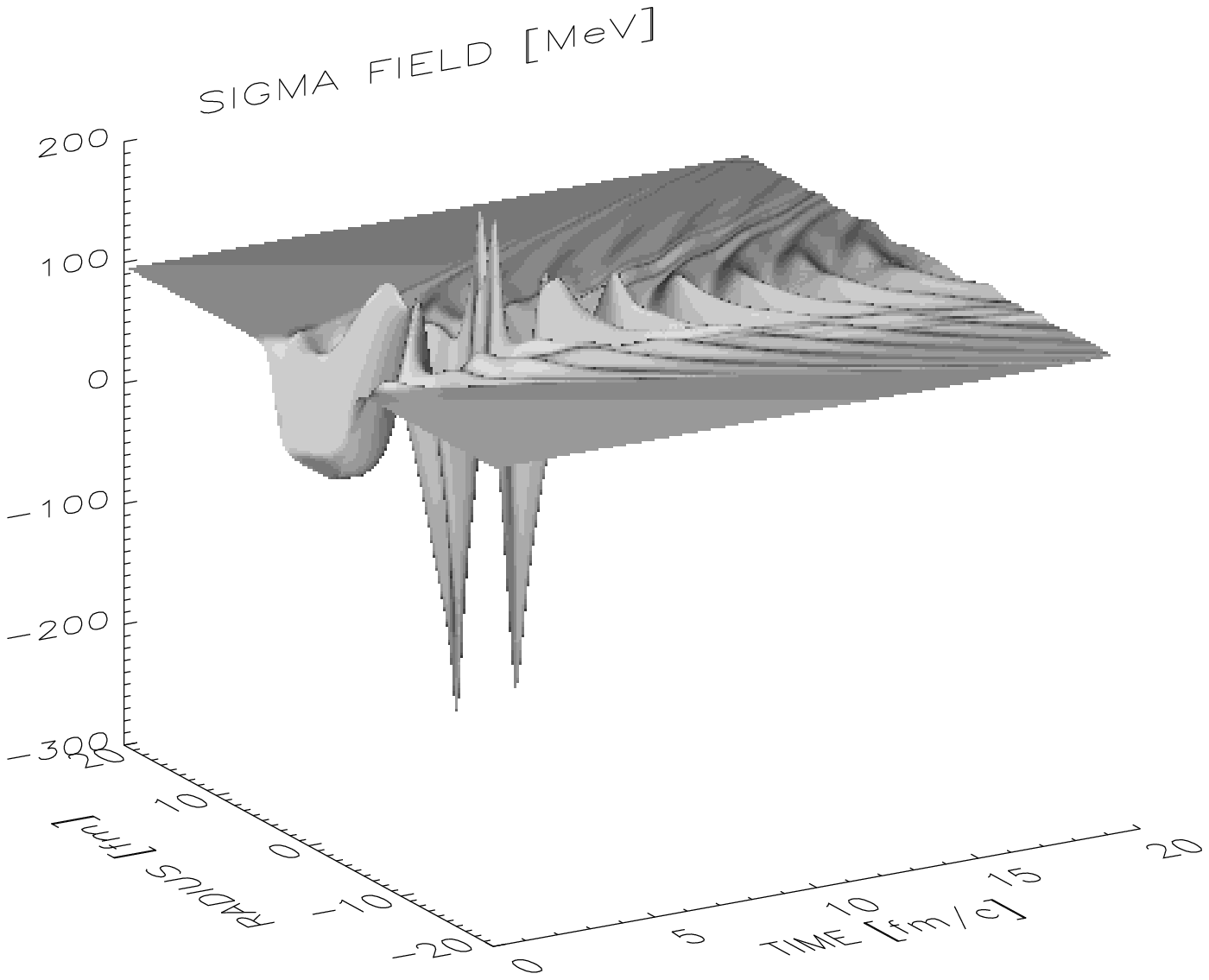,height=7.5cm}}}
\caption{The $\pi_0$ and $\sigma$ field strengths within the forward light-cone 
for $g=5.5$ and initial condition $\Phi=(0.1,0.07,0,0)f_\pi$,
$\partial \Phi/\partial t=0$. The initial expansion scalar was chosen as
$\partial\cdot u\approx1/(2{\rm fm/c})$.}
\label{fields}
\end{figure}
\begin{figure}[htp]
\centerline{\hbox{\epsfig{figure=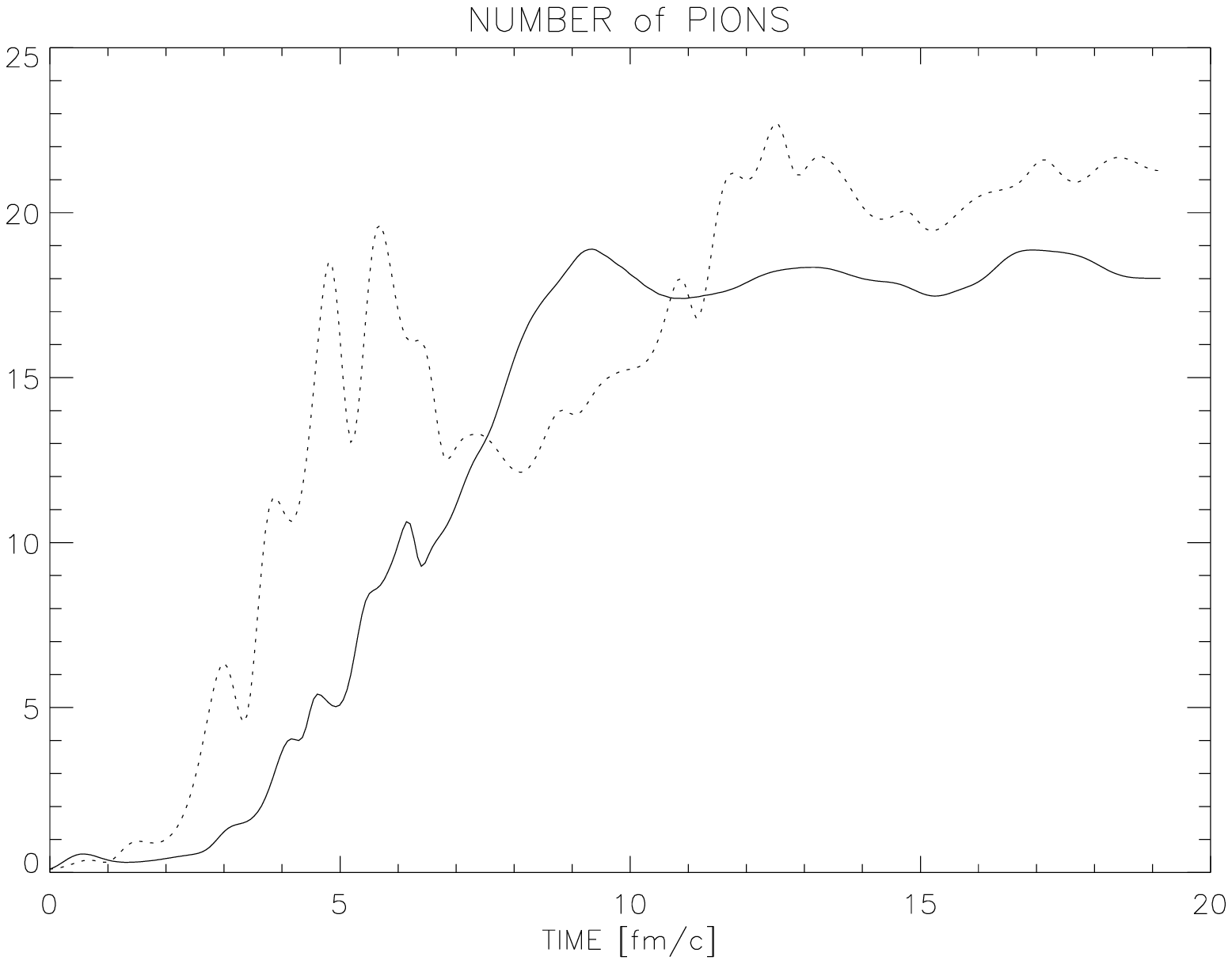,height=6.5cm} \hspace{-1cm}
\epsfig{figure=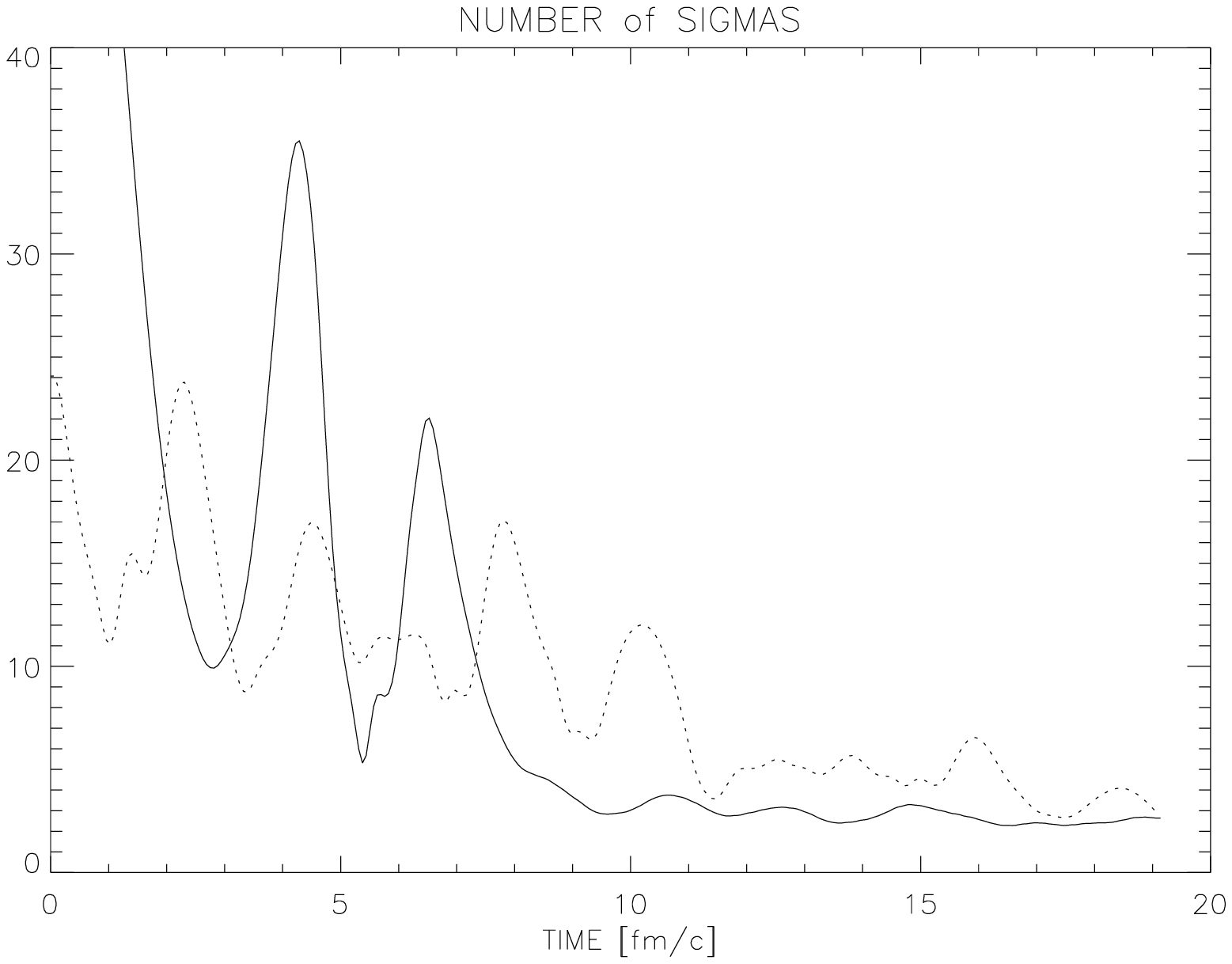,height=6.5cm}}}
\caption{Number of $\sigma$ and $\pi_0$ mesons produced from the
decay of the classical chiral field. The solid lines correspond to 
the self-consistent calculation with first order phase transition (initial
conditions as in Fig. \protect\ref{fields}); dotted lines
are for the ``quenched'' initial conditions $\Phi=(0.45,0.07,0,0)f_{\pi}$,
$\partial \Phi/\partial t=0$, $V\equiv U$ (this corresponds to roughly 
the same potential energy at the time when ``roll-down'' begins).}
\label{num}
\end{figure}

  
Let us consider the evolution starting close to the minimum of the
 potential shown to the
left in Fig.~\ref{potstrong}. The evolution of the fields is shown in
Fig.~\ref{fields}. The chiral field is ``trapped'' for $t\sim2$~fm/c in the 
local minimum 
until the temperature drops to the value corresponding to the potential to 
the right ($T\sim100$~MeV). 
At this point the bubble is supercooled by about 15\% and the barrier to
the global minimum has almost disappeared.
Here, the field starts to ``accelerate'' and rolls towards the true
vacuum. Note that the field gets an additional ``kick'' because the bubble
starts to collapse under the vacuum pressure~\cite{Ove}, which at this point,
after the
system has crossed the first order phase transition line, exceeds 
that of the classical field inside the bubble. For this scenario, we find
 that 
the pion field is amplified by more than a factor of 100, corresponding to
 about 18 $\pi_0$'s and 3 $\sigma$'s for an initial radius of 4~fm,
 see Fig. \ref{num}. For the simple quench scenario, where the fields
 evolve in the
zero temperature potential $U$ and without coupling to any heat-bath,
we get similar results: 21 $\pi_0$'s
and 4 $\sigma$'s.  We computed these numbers as described
 in~\cite{Amelino,Ove}. 

In summary, we discussed a novel mechanism for DCC formation in a first-order 
chiral phase transition. 
We study a bubble of disoriented chiral field that decouples from the rest of
the system before reaching the phase boundary.
The bubble supercools and grows 
due to the large volume-expansion rate established during preexpansion
(before decoupling of the bubble).
As the effective potential approaches the $T=0$ form, the local minimum at
$\Phi=0$ disappears. 
This leads to the ``quenched'' initial condition.
The subsequent alignment in the vacuum direction
leads to very strong amplification of low momentum modes of the pion field. 
Numerical simulations within the linear $\sigma$-model coupled to
quarks at non-zero baryon-chemical potential are in progress.
However, the above general discussion 
does not depend on a specific model for the chiral dynamics.

If the chiral phase transition is first-order, as is particularly
likely the case for finite
baryon density~\cite{Stephanov:1999zu}, and if the system breaks up
into smaller droplets or bubbles, rapidity fluctuations
(e.g.\ of baryon number) can occur, as proposed 
in~\cite{andy,Mishustin,SG2,Csernai:1995zn}.
If a DCC is formed,
in coincidence ``pion-spikes'' of given bubble-isospin could
appear in the same $p_T$ and rapidity range.
\acknowledgements
We thank the Yale Relativistic Heavy Ion Group for kind hospitality and
support from grant no.\ DE-FG02-91ER-40609.
Also, we gratefully acknowledge fruitful discussions with A.D.\ Jackson,
I.\ Mishustin, D.H.\ Rischke, J.\ Schaffner and U.A.\ Wiedemann.
We thank D.H.\ Rischke for reading the manuscript prior to publication.

\end{document}